# Black phosphorus nanodevices at terahertz frequencies: photodetectors and future challenges

Leonardo Viti,[1] Antonio Politano,[2] Miriam Serena Vitiello[1]

[1]*NEST, Istituto Nanoscienze – CNR and Scuola Normale Superiore, Piazza San Silvestro 12, Pisa, I-56127, Italy.*

[2] *Dipartimento di Fisica, Università degli Studi della Calabria, via Ponte Bucci , 87036 Rende (CS) , Italy.*

The discovery of graphene triggered a rapid rise of unexplored two-dimensional materials and heterostructures having optoelectronic and photonics properties that can be tailored on the nanoscale. Among these materials, black phosphorus (BP) has attracted a remarkable interest thanks to many favorable properties, such as high carrier mobility, in-plane anisotropy, the possibility to alter its transport via electrical gating and direct band-gap, that can be tuned by thickness from 0.3 eV (bulk crystalline) to 1.7 eV (single atomic layer). When integrated in a microscopic field effect transistor (FET), a few-layer BP flake can detect Terahertz (THz) frequency radiation. Remarkably, the in-plane crystalline anisotropy can be exploited to tailor the mechanisms that dominate the photoresponse; a BP-based field effect transistor can be engineered to act as a plasma-wave rectifier, a thermoelectric sensor or a thermal bolometer. Here we present a review on recent research on BP detectors operating from 0.26 THz to 3.4 THz with particular emphasis on the underlying physical mechanisms and the future challenges that are yet to be addressed for making BP the active core of stable and reliable optical and electronic technologies.

## 1. INTRODUCTION

The rising interest in Terahertz (THz) radiation (loosely defined as the 0.1−10 THz frequency range, 30 - 3000 μm wavelength range) has been triggered in the last decade by a wealth of applications in security, biomedical imaging, gas sensing, non-destructive testing and materials analysis, non-contact imaging of coatings and composites, non-invasive medical diagnosis of tumors and dental diseases [1]. However, the full exploitation of this underdeveloped portion of the electromagnetic spectrum requires the development of sensitive and performing systems exploiting powerful, stable and coherent sources as well as fast, sensitive and portable photodetectors. These key priorities prompted a major surge of interdisciplinary research for the investigation of different technologies in-between optics and microwave electronics and a large variety of material systems offering *ad-hoc* properties to target the needed performance and functionalities.

In this context, several room-temperature (RT), on-chip integrated, detection technologies, suitable for real-time frame acquisition and based on distinctive physical mechanisms have been recently proposed and implemented [2]. Amongst them, the present most common architectures rely on semiconductor micro-bolometers [3], fast non-linear rectifying electronics such as Shottky diodes [4], high electron mobility transistors (HEMTs) and field effect transistors (FETs) [5, 6]. The latter can be realized with standard complementary metal-oxide semiconductor (CMOS) or silicon technology and have already shown potential for the development of performing and cost effective THz detection systems [7].

Two-dimensional (2D) layered materials, such as graphene [8, 9], BP [10, 11] or transition metal dichalcogenides (TMDs) [12], display an exceptional technological potential for devising flexible photodetectors that can operate over the broad frequency range from the visible to the THz.

Being compatible with many substrates, they can be easily integrated in an array configuration, and, exceptionally, can allow to engineer the physical mechanisms governing light detection *from scratch* [13], reaching high electrical bandwidths and low noise. Here we review the detection performances of BP-based nano-devices at THz frequencies.

## 2. BLACK PHOSPHORUS

The development of mechanical exfoliation techniques prompted extensive research on layered 2D materials, aiming to the study of their unique physical properties and widening their scope of application [14]. Among these materials, graphene is appealing for its unmatched optical properties, stemming from the absence of an energy gap, which enables charge carrier generation by light absorption over a very wide energy spectrum. Furthermore, graphene exhibits ultrafast carrier dynamics, tunable optical properties via electrostatic doping and high-mobility (up to $10^5$ $cm^2V^{-1}s^{-1}$), thanks to the linear energy-momentum dispersion of the band structure near the Dirac point [13]. However, the lack of an energy gap, a kind of inherent "leakage" for carrier transport, limits the performance of graphene-based electronic devices as photodetectors or digital logic circuits, where low dark currents and large conductivity switching are respectively required.

At the same time, the field of 2D layered crystals has recently been populated with other materials, having complementary features with respect to graphene. For example, TMDs ($MoS_2$, $MoSe_2$, $WS_2$, $WSe_2$, etc.) show a direct, finite band gap (0.4 – 2.3 eV) which endows them with a very high carrier density tuneability (on/off current ratios in a transistor up to $10^{10}$), but limits the carrier mobility to relatively low values, typically less than 250 $cm^2V^{-1}s^{-1}$[15].

In this context, black phosphorus (BP), the most stable allotrope of the element phosphorus, appears as a natural trade-off between graphene and TMDs. Indeed, with its thickness dependent energy gap, spanning from 0.3 eV in the bulk case, to 1.7 eV of the monolayer case (phosphorene) [16], BP can reach on/off ratio up to $10^5$ [17] in a FET, still retaining a room-temperature mobility above 1000 $cm^2V^{-1}s^{-1}$[18], suitable for the development of RF speed (~ 1 GHz) switching circuits [19]. Moreover, the interband optical transitions cover the spectral range from the visible to the mid-infrared, allowing for its application as a fast receiver in the telecom band [10].

Differently from graphene, the atomic planes of BP, bound together by van der Waals forces, do not present an atomically flat structure [20]. This so-called puckered honeycomb arrangement of atoms depicted in Figure 1 defines two in-plane crystalline directions: the armchair axis (or x-axis) and the zigzag axis (or y-axis), respectively orthogonal and parallel to the corrugation direction. This peculiarity provides BP with an inherent anisotropy that involves its optical, electrical and thermal properties. Yuan H. *et al.* [21] recently demonstrated that the anisotropic band structure entails a broadband linear dichroism: when a radiation with energy large enough to excite optical transitions (i.e. larger than the band gap) impinges on BP, it experiences an absorption coefficient that can be controlled by changing the light polarization state.

The anisotropic band dispersion also affects the effective masses of free carriers in BP: the effective mass along the armchair direction is about an order of magnitude smaller than that along the zigzag direction [22]. Hence, the carrier mobility (μ) and electrical conductivity (σ) varies significantly between the two crystalline orientations, $σ_x$ being typically two times larger than $σ_y$ [23]. Conversely, the thermal conductivity (κ) is larger along the zigzag direction. This is due to the different arrangement of atomic sites, which is more robust to longitudinal vibrations along the y direction, making the longitudinal in-plane acoustic (LA) phonon speed (8397 m/s) larger with respect to that occurring in the armchair direction (4246 m/s) [22]. Since κ is proportional to the square of the speed of sound, the material thermal conductivity along the y-axis ($κ_y$) is larger than its thermal conductivity along the x-axis ($κ_x$). Therefore, the direction of maximum thermal conductance lies perpendicularly to the direction of maximum electrical conductance.

A suitable method to determine the crystalline quality and in-plane orientation of an exfoliated few-layer BP flake (Figure 1b) is micro-Raman spectroscopy. Indeed, Raman spectra present three characteristic peaks at 362, 440, and 468 cm$^{-1}$ (Figure 1c) corresponding to the $A_g^1$, $B_{2g}$, and $A_g^2$ phonon modes, whose relative intensities change if the incident light is polarized along the x-axis, the y-axis, or at a 45° angle between the two axis (this direction is typically labeled as D-axis) [23].

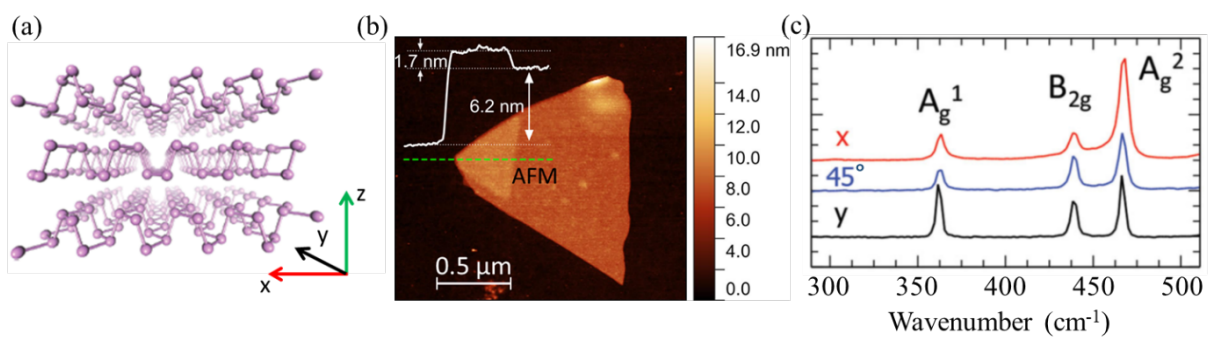

FIG. 1. (a) BP atoms are arranged in puckered honeycomb layers bounded together by van der Waals forces; the armchair (x) and zigzag (y) crystal axis are shown on the graph. Reprinted by permission from Macmillan publishers Ltd.: *Nature Nanotechnology* (ref. [20]), © 2014. (b) Atomic force microscopy (AFM) topographic image of an individual flake having a thickness of 6.2 nm. The related topographic line profile, acquired along the dashed green line, is shown on the graph. Image taken from ref. [11] with permission of John Wiley & Sons, Inc. (c) Micro-Raman spectra collected by exciting BP flakes along the z-axis with the 532 nm line of an Nd-YAG pumping laser and by varying the polarization between the armchair (x), the zigzag (y), and the 45° orientation angle: peaks are found at 362, 440, and 468 cm$^{-1}$, corresponding to the $A_g^1$, $B_{2g}$, and $A_g^2$ vibrational modes, respectively. Reprinted from ref. [11] with permission of John Wiley & Sons, Inc.

Owing to this unique set of physical properties, black phosphorus promises new functionalities over a variety of device architectures, like sources, modulators, passive components and photodetectors. For example, owing to its ultrafast nonlinear optical response, it has been proposed as a saturable absorber in the telecom band (1400 nm -1600 nm) for the realization of passively mode-locked lasers [24].

Recently, Huber *et al.* [25] revealed the ultrafast (~50 fs) switching dynamics of interface polaritons in SiO$_2$/BP/SiO$_2$ heterostructrures. The spectral purity and coherence of the activated phonon-plasmon-polariton modes can provide a versatile and robust technological platform polariton-based mid-infrared optoelectronic devices and, remarkably, for ultrafast plasmonic applications.

Moreover, thanks to the inherent in-plane anisotropy, BP has been exploited to devise high-speed photodetectors with polarization sensitivity in the visible and near-infrared range [10, 26]. Recent works from our group [11, 27] have also demonstrated the first implementations of THz detectors realized with antenna-coupled FETs based on BP flakes, where we took advantage of the in-plane material anisotropy, combined with specific coupling designs, to trigger different physical dynamics to mediate light detection.

### 3. TERAHERTZ DETECTION MECHANISMS IN BP-BASED FETs

The basic principle of photodetection is the conversion of absorbed photons into an electrical signal. Unlike the case of visible light or near-infrared radiation, the energy carried by a THz photon (hν = 4 meV for ν = 1THz) is not large enough to excite an optical transition across the energy gap of BP, i.e. does not create extra free carriers. For this reason, the generation of an electrical signal induced by the incoming radiation must rely either on electronic or thermal effects.

The integration of a BP flake within a FET can trigger both dynamics when light is impinging on it. Under this configuration, three main effects can indeed take place in the device: the over-damped or resonant plasma-wave, the thermoelectric and the bolometric effects [27].

The activation of the plasma-wave mechanism in a FET detector can be interpreted as deriving from the non-linear dependence of the FET channel current on the gate voltage ($V_G$) near the pinch-off point. These devices have the advantage that the responsivity can be maximized with $V_G$, while measuring the output at the drain with no source-drain bias applied, thus dramatically reducing the noise caused by a *dc* current. This physical effect, predicted by Dyakonov and Shur [28], results in a *dc* voltage generated along the channel of a FET in response to the oscillating electromagnetic field applied, in a strongly anti-symmetric architecture, between the source (S) and gate (G) electrodes.

In order to funnel the free space THz radiation in the deeply sub-wavelength active element, a proper light harvesting scheme is needed. The typical approach is the exploitation of planar dipole antennas whose arms are connected to the S and top-gate electrodes (Figure 2a). This allows the *ac* field carried by the THz wave to be asymmetrically fed into the transistor channel, simultaneously modulating the density and drift velocity of carriers. This leads to a rectified current proportional to the square amplitude of the field itself, hence to the intensity of the incoming radiation. At RT, the collective density oscillations (plasma-waves), launched from the source side of the FET, are overdamped, i.e. they decay before reaching the drain (D) contact. In this regime, the detection is inherently broadband and can be well described by a distributed resistive self-mixing scheme [29]. According to the theoretical models, the amplitude of the photoresponse ($\Delta u_p$ in the case of a photovoltage) is expected to vary as a function of the applied gate voltage ($V_G$) and can be retrieved from the *dc* transfer characteristics of the FET via the relation [30]:

$$\Delta u_p = \frac{U_a^2}{\sigma}\frac{d\sigma}{dV_G} \cdot \left[\frac{R_L}{\frac{1}{\sigma} + R_L}\right] \cdot \eta \qquad (1)$$

where $U_a$ is the amplitude of the *ac* radiation field, $R_L$ is the finite impedance of the measurement setup including the readout circuitry, and η is a proportionality factor that takes into account the coupling efficiency of the THz antenna.

The asymmetric antenna design generates a field enhancement between the G and S contacts of the FET. With reference to Figure 2a, this can also result in a temperature difference between the source and the drain side of the transistor. Therefore, under THz illumination, a thermal gradient can arise along the channel, leading to the diffusion of carriers from the hot to the cold regions. This phenomenon is the basis for the thermoelectric effect; if a thermal difference (ΔT) is applied along a material with non-zero Seebeck coefficient ($S_B$), electrically connected by two metallic leads, a voltage difference $V_T$ develops between the two contacts and $V_T = S_B \cdot \Delta T$. Since BP is typically a p-type degenerate semiconductor at RT, the value of $S_B$ is positive and depends on the gate voltage following the Mott relation:

$$S_B = \frac{\pi^2 k_B^2 T}{3e} \frac{1}{\sigma} \frac{d\sigma}{dE}\bigg|_{E=E_F} \quad (2)$$

Where $k_B$ is the Boltzmann constant, $e$ is the (positive) elementary charge, σ(E) is the conductivity as a function of energy and the derivative is evaluated at the Fermi energy $E_F$. At room temperature $S_B$ values as large as 335 µV/K have been found in few-layer BP flakes [31]. Therefore, if THz radiation impinges on an *asymmetric* antenna structure, both the thermoelectric and plasma wave effects can take place.

However, it is still possible to select the active mechanism by exploiting the relative orientation of the antenna and the crystal directions. Indeed, the thermoelectric effect is maximized when the thermal gradient along the FET channel is maximum. This is achieved by placing the S-D axis parallel to the armchair direction, where κ is at its minimum and σ is at its maximum value. In this configuration, the lack of energetic balance between the hot-side and cold-side of the channel is mainly compensated by the carrier diffusion, resulting in a large thermoelectric voltage that typically overshadows the plasma-wave effect [11]. By choosing a different orientation, for example at 45° between the x and y directions (D axis), the thermoelectric contribution can be reduced and the plasma-wave mechanism is observed [27].

On the other hand, in order to activate selectively the bolometric effect, a *symmetric* antenna structure is needed. In fact, this mechanism arises as a consequence of the uniform heating of the FET channel when the antenna funnels the impinging THz beam. This can be achieved in a totally symmetric device architecture, like the one shown in Figure 2b, where the plasma-wave and thermoelectric effects would be inhibited. The amplitude of the bolometric photovoltage ($V_b$) is directly related to the temperature dependence of the conductivity via the relation [27]:

$$V_b \propto \frac{1}{\sigma} \cdot \frac{\partial \sigma}{\partial T} \quad (3)$$

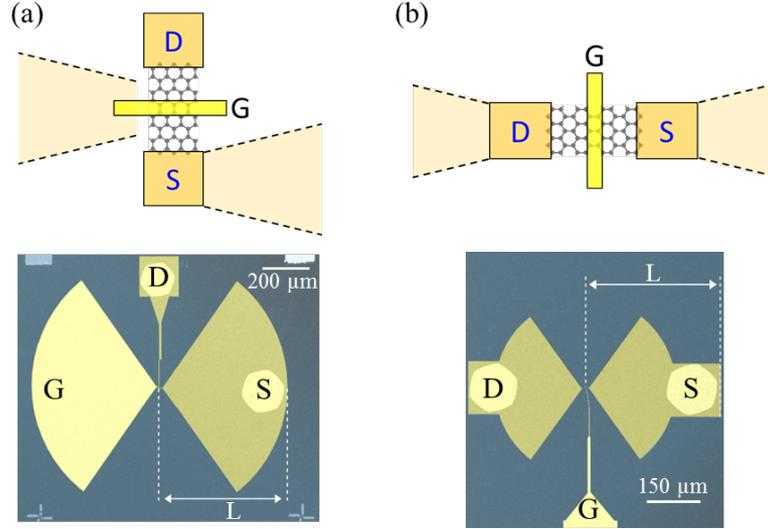

FIG. 2. (a) An asymmetric dipole antenna configuration is lithographically designed by connecting the two arms of the dipole to the S and G electrodes. The false colors scanning electron microscope (SEM) image shows an asymmetric bow-tie antenna with dipole arm length L = 500 μm (taken from [27]). (b) In a symmetric antenna the two-dipole arms are electrically connected to the S and D electrodes. In this configuration S and D are interchangeable. The false colors SEM image shows a symmetric bow-tie antenna with arm length L = 250 μm (taken from [27]).

## 4. SUB-TERAHERTZ DETECTION

The activation of these three detection effects in BP-based FETs has been recently engineered and demonstrated in the sub-THz frequency range (265 GHz – 640 GHz) [11, 27, 32]. The fabrication of BP-FETs begins with the mechanical transfer of thin flakes (~ 8 – 15 nm) on an insulating substrate under an inert atmosphere. In Refs. [11, 27] single flakes were individually contacted with S and D electrodes consisting of a proper adhesion layer/metal sequence (Cr/Au [11] or Ni/Au [27]). The shape of the electrodes was defined by aligned electron beam lithography (EBL). A $SiO_2$ layer was then deposited on the sample via Ar sputtering, so that the exposed face of the BP-flake is fully encapsulated within it. This step prevents degradation due to the chemical reaction of BP with oxygen, which is known to modify the electrical device performances over time and deteriorate the flake itself [33]. The top-gate electrode was aligned with the center of the channel via EBL and defined via thermal evaporation of an 80 nm layer of Cr/Au.

The dominant detection dynamics mainly depends on two device parameters: (*i*) the symmetric (or asymmetric) antenna design and (*ii*) the relative orientation between the source-drain axis and the crystalline in-plane direction.

Three different combinations of antenna design and crystal orientation with respect to the channel direction were tested. Sample A (Figure 3a) was equipped with an asymmetric planar bow-tie antenna (total dipole length 500 μm) with the S-D axis oriented along the armchair direction of the BP flake. The total channel length was 2.7 μm and the gate width was 580 nm. The estimated room temperature field effect mobility ($\mu_{FE}$) for this device was $\mu_{FE}$ = 470 $cm^2V^{-1}s^{-1}$. Sample B (Figure 3c) was equipped with an asymmetric 1 mm long bow-tie antenna with the S-D axis oriented at 45° with respect to the armchair direction. The total channel length was 1.8 μm and the gate width was 1.0 μm. For this device a mobility $\mu_{FE} = 330$ cm$^2$V$^{-1}$s$^{-1}$ was obtained. Sample C (Figure 3e) was equipped with a symmetric 500 μm long bow-tie antenna with the S-D axis oriented along the zigzag direction. The total channel length was 900 nm and the gate width was 450 nm. For this device a mobility $\mu_{FE} = 380$ cm$^2$V$^{-1}$s$^{-1}$ was found.

Experiments were carried out at room temperature with a ~ 300 GHz, linearly polarized, radiation produced by a tunable electronic source. The THz beam, modulated by a mechanical chopper, was focused on the detector by a set of off-axis parabolic mirrors, reaching a spot diameter of 4 mm in the focal point. The photoresponse was measured in photovoltage mode [27]: the S contact was kept grounded and a *dc* voltage (*LIA*) was measured at the D electrode with a lock-in amplifier. The amplitude of *LIA* was proportional to the incoming THz intensity and the detector responsivity ($R_v$) was varied by changing the gate bias. $R_v$, defined as the ratio between the output signal and the input optical power, is evaluated using the following expression [11]:

$$R_v = \frac{\sqrt{2}\frac{\pi}{2} \text{LIA}}{G_n P_{\text{tot}}} \cdot \frac{S_t}{S_\lambda} \tag{4}$$

Where the factor $\pi/\sqrt{2}$ is a normalization coefficient that takes into account the *rms* value of the fundamental sine wave Fourier component of the square wave produced by the chopper, $G_n$ is the gain of the preamplifier, $P_{\text{tot}}$ is the average electromagnetic power, $S_t$ is the beam spot area and $S_\lambda$ is the diffraction limited area ($S_\lambda = \lambda^2/4$).

Figure 3b shows $R_v$ as a function of $V_G$ for Sample A. The comparison with the dependence of the Seebeck coefficient with respect to $V_G$ [11], proportional to the predicted photo-thermoelectric response $V_T$, shows a good qualitative agreement, demonstrating the activation of the thermoelectric effect.

On the other hand, as depicted in Fig. 3d, the $R_v$ vs. $V_G$ plot for sample B, obtained by shining a 294 GHz radiation, exactly follows the dependence of the photoresponse $\Delta u_p$ predicted by the Dyakonov-Shur theory (Fig. 3d). This difference between samples A and B is caused by the different relative orientation between the S-D axis and the crystalline in-plane direction in the two cases.

When a symmetric THz antenna is employed (sample C) the bolometric effect is expected to occur. The responsivity curve of sample C is shown in Fig. 3f. An immediate difference with respect to the results obtained for samples A and B is the sign change in $R_v$ for $V_G = -1.1$ V. Interestingly, the same sign change is also found in the bolometric coefficient [27]. The direct comparison between the $R_v$ vs. $V_G$ and the $\sigma^{-1} \cdot \partial\sigma/\partial T$ vs. $V_G$ curves shown in Fig. 3f, demonstrates that sample C behaves like a THz bolometer.

Amongst the three reported devices, a maximum responsivity value $R_v = 7.8$ V/W has been obtained for sample C, corresponding to a noise equivalent power (NEP, defined as the ratio between $R_v$ and the noise spectral density) of 7 nW/√Hz [27].

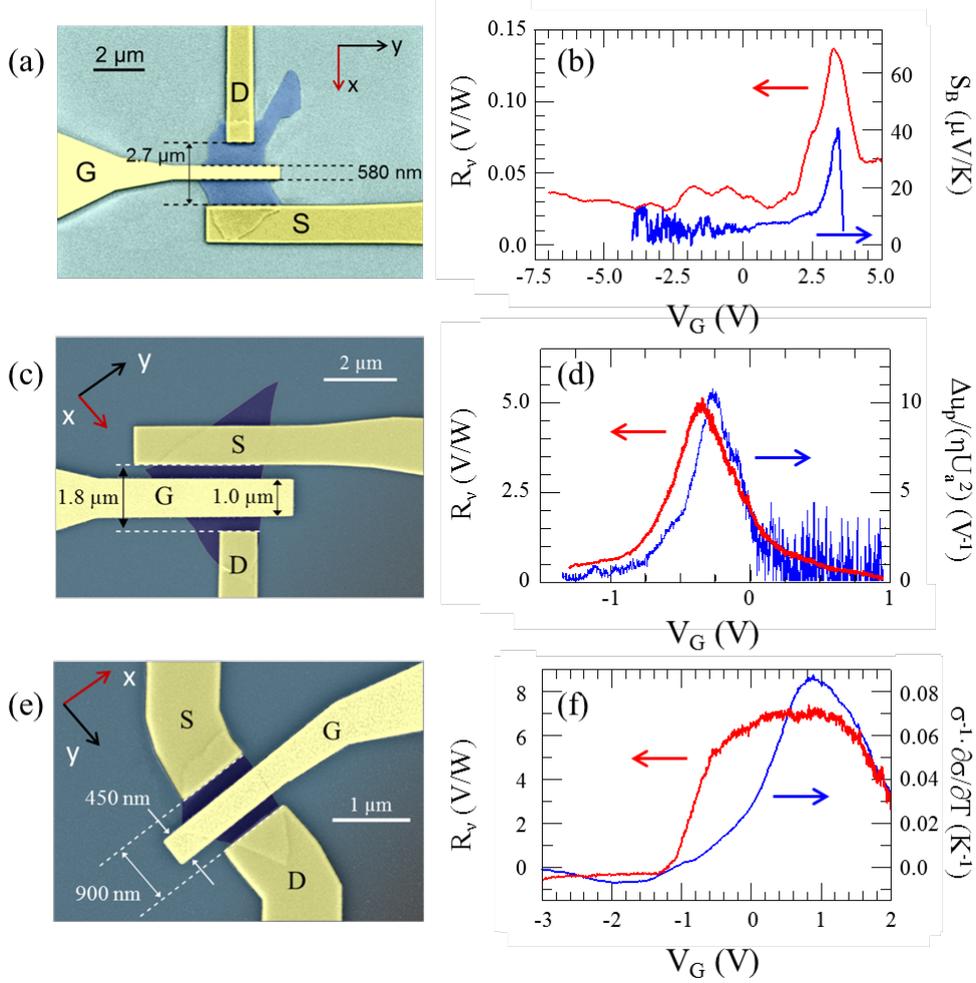

FIG. 3. Sub-THz responsivity. (a,b) Characterization of Sample A: the antenna is asymmetric and the S-D axis is oriented along the x direction. (a) False colors SEM image of sample A. The channel length is 2.7 μm and the gate width is 580 nm. Reprinted from ref. [11] with permission of John Wiley & Sons, Inc. (b) Left vertical axis: gate bias dependence of the RT responsivity. Right vertical axis: Seebeck coefficient extracted from the measured transfer characteristics of the FET (equation 2). (c,d) Characterization of Sample B: the antenna is asymmetric and the S-D axis is oriented along the D direction. (c) False colors SEM image of sample B (taken from [27]). The channel length is 1.8 μm and the gate width is 1.0 μm. (d) Left vertical axis: gate bias dependence of the RT responsivity. Right vertical axis: predicted photoresponse trend as a function of $V_G$ under the overdamped plasma-wave regime, calculated using equation 1. (e,f) Characterization of Sample C: the antenna is symmetric and the S-D axis is oriented along the y direction. (e) False colors SEM image of sample C (taken from [27]). The channel length is 900 nm and the gate width is 450 nm. (d) Left vertical axis: gate bias dependence of the RT responsivity. Right vertical axis: predicted bolometric trend, estimated from the gate voltage dependence of the quantity $\sigma^{-1} \cdot \partial\sigma/\partial T$ (equation 3).

These pivotal experiments have underlined the potentiality of BP-FETs to be engineered in view of activating specific physical mechanisms, pushing a further research on the topic. Therefore, more recently, the realization of detectors operating in the 0.3-0.65 THz range with a combination of hexagonal boron nitride (hBN) and BP flakes has been reported [32]. Owing to the flatness of hBN and to its compatibility with the honeycomb lattice, the encapsulation of BP in hBN/BP/hBN van der Waals heterostructure leads to extremely inert and air stable devices [32, 34]. Furthermore, with this approach, record mobility values have been reached (up to ≈1350 cm$^2$ V$^{-1}$ s$^{-1}$ at room temperature [18] and ≈ 6000 cm$^2$ V$^{-1}$ s$^{-1}$ for T < 30 K [35]).

THz detectors realized with hBN/BP/hBN stacks have also a higher thermal stability with respect to oxide-encapsulated devices. The reported stability, allowed performing low temperature experiments which revealed and a signal-to-noise ratio (SNR) of $2 \times 10^4$ at 4 K [32]. Importantly, the selective activation of the three detection mechanisms was demonstrated in a single BP-FET, equipped with an asymmetric antenna and oriented along the D axis. For T < 50 K the detection was dominated by the bolometric effect, whereas at RT there was a simultaneous coexistence of the thermoelectric and plasma-wave effects, whose relative amplitudes were modified by changing the frequency of the incoming radiation [32].

## 5. DETECTION AT 3.4 THz

Many interesting applications in the THz frequency range require the use of compact, coherent, broadly tunable, monolithic micro-sources, having high spectral purity. Quantum cascade lasers (QCLs) are certainly the most appropriate in this respect: they operate in the 1.2-5.4 THz frequency range, with peak output powers that can reach watt-level, continuous-wave operation and ultra-narrow linewidth, which make them versatile for applications in high-resolution spectroscopy, gas sensing, and heterodyne detection [36]. To exploit the potential of QCLs, suitable detectors operating at frequencies > 1THz are needed.

FETs can provide efficient THz rectification far beyond their transit-time limited cut-off frequency: Si MOSFETs operating up to 4.3 THz [37] and nanowire-based FET operating up to 2.8 THz have been indeed demonstrated in the last years [38]. In those devices, the transition between a sub-THz to a few-THz detection does not affect the active element (i.e. the rectifier), but only the coupling architecture: the antenna dimensions shrink to match shorter wavelengths.

Following this approach, the operation range of a BP-FET can be shifted to higher frequencies by reducing the antenna dimensions, while maintaining the same impedance (~ 70 $\Omega$ for a bow-tie). To match the wavelength generated by a 3.4 THz QCL, a BP-FET equipped with a 32µm long split-bow-tie antenna was fabricated (Fig. 4a). The channel length was set to 900 nm, while the width of the G electrode was 450 nm. The S-D axis was oriented at 45° with respect to the armchair crystalline direction, hence the plasma-wave rectification effect is expected to dominate the detection.

The light beam emitted by the QCL was focused on the detector using a pair of 5 cm focal length Picarin lenses, obtaining a Gaussian shape with ~ 200 µm diameter (Fig. 4b). The device was tested in a photovoltage-mode configuration: the S electrode was grounded, $V_G$ was set with a *dc* voltage generator and the response was measured at the D electrode with a lock-in amplifier. A voltage preamplifier with gain $G_n$ = 40 dB was used in the experiments. Figure 4c shows the dependence of $R_v$, evaluated by using equation 4, as a function of the applied $V_G$: a maximum peak of 4.5 V/W was obtained for $V_G$ = 0.65 V.

The predicted plasma-wave response for this sample, calculated from the static *dc* conductivity as $\Delta u_p/(\eta U_a^2)$ (see eq.1), is depicted in Fig. 4d. A maximum is expected for $V_G$ = 0.6 V, whereas for $V_G$ < 0.3 V the response is predicted to drop rapidly. The qualitative agreement between the two graphs of Fig. 4c and 4d, confirms that in a sample with a 45°-oriented asymmetric antenna, the response is dominated by the overdamped plasma-wave contribution even at 3.4 THz. This broadband behavior is in agreement with the inherently broadband nature of the overdamped plasma-wave mechanism.

Figure 4e shows the NEP of the device, calculated assuming the noise spectral density to be dominated by the Johnson-Nyquist noise: $N_{th} = (4R_{ch} \cdot k_B T)^{1/2}$, where $R_{ch}$ is the channel resistance and $k_B$ is the Boltzmann constant. A minimum NEP of 5 nW/Hz$^{1/2}$ was found for $V_G$ = 0.65 V; this value is lower than the NEP obtained at RT for detectors operating at sub-THz frequencies, proving the efficient scalability of the BP-FET device concept over a broad frequency range.

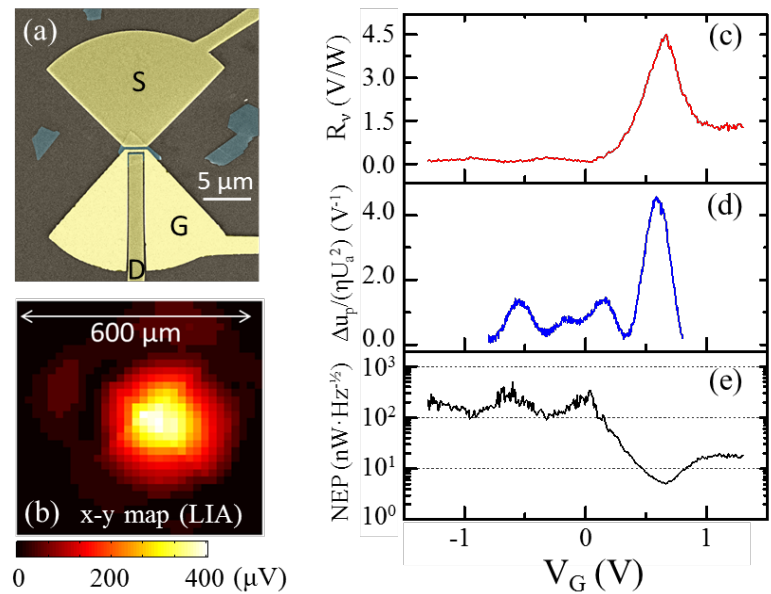

FIG. 4. Detection at 3.4 THz. (a) False colors SEM image of the top-gated BP-FET. The asymmetric split-bow-tie antenna has a total length of 32 μm. The channel length is 900 nm, the gate width is 450 nm and the S-D axis is oriented along the D direction. (b) THz image of the focused QCL beam, obtained by scanning the detector on the focal plane, while keeping $V_G$ = 0.65 V. The beam spot radius is ~ 100 μm. (c) $R_v$ vs. $V_G$ dependence. (d) Predicted photoresponse trend as a function of $V_G$ under the overdamped plasma-wave regime, calculated using equation 1. (e) NEP estimated from the ratio between $R_v$ and the Johnson-Nyquist noise spectral density.

## 6. PERSPECTIVE AND CHALLENGES

The recent progress in the development of photodetectors [10, 11], fast electronic and photonic devices [19, 24, 25], electrode materials in lithium and sodium ion batteries [39], supercapacitors [40] and quantum dots [41], is going to speed up the practical future implementation of BP atomic layers in optoelectronics and nano-optics. Notably, the material shows a great potential for applications where excellent switching speed is required.

Of particular interest for photonics is the exploration of plasmon polaritons - hybrid light matter modes involving the collective oscillation of mobile charges. In graphene, plasmon polaritons have been exploited to confine and manipulate infrared radiation to length scales of tens of nanometres. The imaging of long-range propagating phonon-plasmon-polaritons and their ultrafast optical switching has been also demonstrated in black phosphorous via near-field optical microscopy experiments (Figure 5). In particular, femtosecond pulses can initiate the plasmonic response in this semiconductor through direct excitation of electrons across the bandgap. The reported switching speed of activated phonon-plasmon-polariton modes in SiO$_2$/BP/SiO$_2$ heterostructures [25] opens an intriguing perspective for the development of ultrafast plasmonic circuits, modulators and switches for a variety of mid-infrared and near infrared nanophotonic applications.

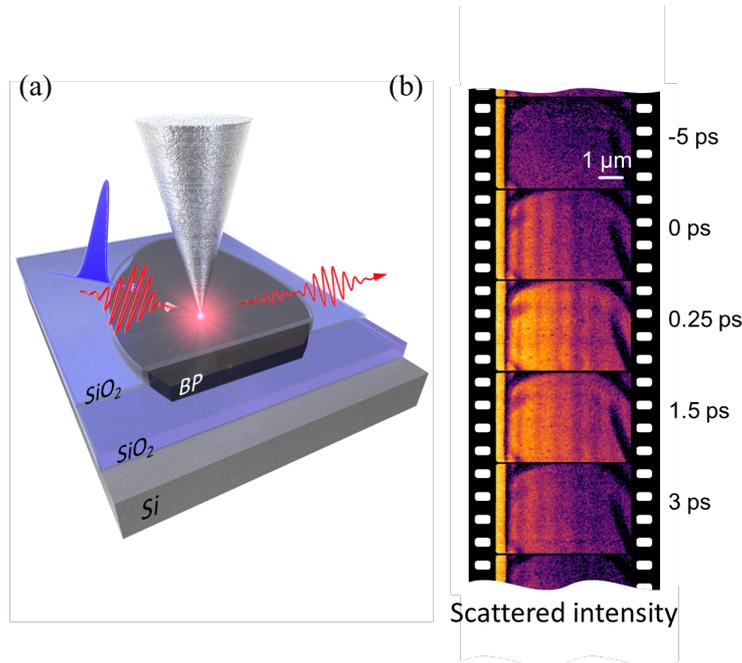

FIG. 5. (a) An ultrafast near infrared (blue) pump pulse is focused onto a heterostructure of $SiO_2$ and BP on a $SiO_2$-capped Si wafer. The sample is probed in the near field with a mid-infrared (red) pulse coupled to a sharp metallic tip. (b) Scattered near-field intensity images of the heterostructure plotted for five different delay times between the pump and probe pulses (scale bar 1 μm). Reprinted by permission from Macmillan publishers Ltd.: *Nature Nanotechnology* (ref. [25]), © 2017.

The evolution of BP-based technologies is however currently hindered by the limited availability of large area flakes and by the fast material reactivity with oxygen, which leads to its quick deterioration in air. The future implementation of electronic and photonic technologies based on this material system should therefore face three major challenges: (*i*) the development of controllable and versatile growth techniques; (*ii*) the improvement of material chemical and electrical properties; (*iii*) the refinement of the fabrication procedures for long-time stable structures.

On the growth side, the recent progress on BP production by mineralizer-assisted gas-phase transformation method provides an efficient alternative to standard high-pressure synthesis or chemical vapor transport for the growth of large-size, high-quality crystals, also allowing the possibility to add dopants during the growth process [42]. However, high-throughput and scalable synthesis methods are still missing, thus hampering the use of BP for large-scale applications.

Furthermore, the optimization of carrier density via chemical doping during growth is of fundamental importance to improve the performances of BP-based devices; for example, selenium doping has been demonstrated to be beneficial for increasing the responsivity of photodetectors operating in the visible range by more than one order of magnitude [42]. Recently, thin flakes of BP doped with tellurium showed a higher mobility (up to 1850 cm$^2$V$^{-1}$s$^{-1}$ at RT) with respect to undoped crystals and, more remarkably, higher stability under ambient exposure [43]. This result provides an interesting solution to the long-standing obstacle of poor environmental stability, circumventing the otherwise mandatory encapsulation of BP via dielectric deposition or mechanical stacking of van der Waals heterostructures, whose fabrication is costly and often demanding.

## 7. CONCLUSIONS

Black phosphorus earned an extraordinary potential among the family of 2D semiconductor materials when recent results unveiled its high carrier mobility, high optical and UV absorption, and other attractive properties, which are of particular interest for optoelectronic and photonic applications.

The development of photodetectors from the ultra-violet to the near-infrared frequency regions has attracted increasing attention owing to the opportunity of band structure engineering via thickness control and to the peculiar absorption properties stemming from the in-plane crystalline anisotropy of BP.

In this brief review, we have provided a summary of recent achievements in the development of BP-based photo-detectors operating in one of the most technologically appealing region of the electromagnetic spectrum: the far infrared. We discussed detection performances in the 0.3-3.4 THz frequency range, unveiling room-temperature responsivity higher than 4V/W and noise equivalent power lower than 10 nW/Hz$^{1/2}$, making this technology highly competitive for practical process and quality control apprlications. We also discussed the influence of the strong in-plane anisotropy of the material for tailoring the detection dynamics, offering an interesting route for engineering device properties "from scratch". Together with the demonstrated plasmonic response of BP heterostrctures, the present review provides a starting point for future studies and design of BP-based THz photonic, plasmonic and nano-electronic systems.

## 8. AKNOWLEDGEMENTS

This work was supported by the European Research Council through ERC grant 681379 (SPRINT), and by the European Union through the program Graphene Core1 - Graphene-based disruptive technologies" Proposal 696656.